\documentstyle[epsf,rotate]{l-aa}

\newcommand{\kms}{\mbox{\,km\,s$^{-1}$}}
\newcommand{\Msun}{\mbox{M$_{\odot}$}}
\newcommand{\chis}{\mbox{$\chi^{2}~$}}
\newcommand{\rchis}{\mbox{$\chi^{2}_{\nu}~$}}

\begin{document}

\title{Infrared spectroscopy of V616 Mon (=A0620--00): the accretion
disc contamination}

\author{T.~Shahbaz, R.M.~Bandyopadhyay and P.A.~Charles}

\offprints{T.~Shahbaz (tsh@astro.ox.ac.uk)}

\institute{Department of Astrophysics, Oxford University, Keble Road,
Oxford, OX1 3RH, UK}

\thesaurus{2(08.02.3; 08.06.3; 08.09.2: V616 Mon (A0620--00): 13.25.5)}

\date{Received ?? ?? 1997 / Accepted ?? ?? 1997}

\maketitle

\markboth{T.~Shahbaz et al: V616 Mon}{}

%******************************************************************************

\begin{abstract}
We have obtained for the first time $K$-band infrared spectra of
the soft X-ray transient V616 Mon (=A0620--00). We determine the 2-$\sigma$
upper limit to the fraction of light
arising from the accretion disc to be 27 percent. The effect
this has on the binary inclination, determined from modelling the
infrared ellipsoidal variations is to increase it by less than 7 degrees
and decrease the mass of the black hole by less than 3.6 \Msun.

\keywords{binaries: general -- stars: fundamental parameters,
stars: individual: V616 Mon (A0620--00) -- X-rays: stars} 
\end{abstract}

%******************************************************************************

\section{Introduction}

The soft X-ray transients (SXTs), sometimes referred to as ``X-ray
novae'', are low-mass X-ray binary systems in which a late-type star
loses material via an accretion disc to a neutron star or black hole.
They undergo brief outbursts, typically lasting a few months with a
recurrence time of the order of decades (see Tanaka \& Shibazaki 1996 and
van Paradijs \& McClintock 1995 for reviews of their X-ray and optical
properties).

The X-ray transient V616 Mon (=A0620--00) was discovered in 1975 (Elvis
et al., 1975). When the system faded into quiescence the binary nature of
the system was established; the orbital period was determined to be 7.75
h and the secondary star was found to be a K dwarf (Oke 1977; McClintock
et al. 1983). McClintock \& Remillard (1986) then went on to measure the
orbital radial velocity curve of the secondary star, thereby obtaining a
firm {\it minimum} mass for the compact object (i.e. the mass function)
of 3.08 \Msun. This result immediately placed V616 Mon amongst the best
black hole candidates (see van Paradijs
\& McClintock 1995). 

High resolution optical spectroscopy can provide further constraints on
the system masses, as a measurement of the rotational broadening of the
secondary star absorption features leads directly to the {\it ratio} of the
component masses (MRW). But the binary inclination, can only be
determined by exploiting the ellipsoidal modulation of the secondary
star, i.e. the variations caused by observing the differing aspects of
the gravitationally distorted star as it orbits the compact object (see
Shahbaz, Naylor \& Charles 1993 and references within). In V616 Mon these
variations have been observed at both optical (McClintock \& Remillard
1986; Haswell et al. 1993) and infrared (IR) wavelengths (Shahbaz, Naylor
\& Charles 1994; hereafter SNC).

SNC determined the binary inclination for V616 Mon by modelling the IR
ellipsoidal modulation of the secondary star. They obtained $i\sim
37^{\circ}$, which when combined with the value for the binary mass
function and mass ratio resulted in a most probable mass for the compact
object of $\sim$10 \Msun (5.1--17.1\Msun; 90 percent confidence).

There are several complications in such a determination of the binary
inclination; inevitably the optical light from the mass donor is diluted
by the flux from the quiescent accretion disc. Also, the ellipsoidal
light curve may be distorted by other variable contributions to the
observed flux; such as the bright spot, star spots and variable disc
emission including the superhump phenomenon. Underestimating the diluting
flux leads to an underestimate in the binary inclination. At optical
wavelengths it is clear that the accretion disc contributes a significant
amount of flux to the optical light curves, which makes any
interpretation of these light curves somewhat dubious. This can be seen
from the scatter and asymmetric maxima observed in the optical light
curves of V616 Mon by McClintock \& Remillard (1986). The veiling by the
continuum emission from the accretion disc has been measured by MRW. They
find that 94 percent or more of the optical flux close to H$\alpha$ comes
from the secondary star, a fraction which rises with increasing
wavelength, while the veiling ratio falls. Therefore one would expect the
veiling by the accretion disc to be very small or even negligible in the
$K$ band.

Haswell et al. (1993) attempted to deduce the binary inclination from
modelling their optical multi-colour light curves with a combination of
an ellipsoidal variation and a grazing eclipse, assuming a large circular
disk. The inclination derived from the grazing eclipse constraint is in
serious disagreement with that determined from the ellipsoidal variations
in the IR light curves (SNC). However, it is known that in low-mass X-ray
binaries and cataclysmic variables the outer, cool regions of the
accretion disc can be a strong source of IR radiation (e.g. Beall et al.
1984; Berriman, Szkody \& Capps 1985). Consequently, this introduces
doubts on the SNC assumption that the observed IR flux has very little
contamination by the accretion disc. This would then imply that any
interpretation of the ellipsoidal variations of the secondary star in the
IR might be misleading. Nevertheless, for the SXT V404 Cyg we have
already shown directly that the IR accretion disc contamination is small;
it only affects the implied mass of the black hole by at most 2 \Msun
(Shahbaz et al., 1996).

We therefore decided to obtain through IR spectroscopy a direct
measurement of any contamination of the IR flux by the accretion disc in
V616 Mon. From this we can determine the effect it has on the
interpretation of the ellipsoidal variation of the secondary star, and
hence on the component masses.

\begin{table*}
\centering
\small{
\caption{Journal of observations}
\begin{tabular}{lccccc}\hline
\hline\noalign{\smallskip}
Object    &  Date       &   UTC   & Airmass  & Exposure time & Comments \\ 
          &             &  (hrs)  & (mags)   & (secs)        & \\ \hline
V616 Mon  &  13/11/1997 & 11:34   & 1.18     & 1280 & $\phi$=0.31  \\
V616 Mon  &  13/11/1997 & 12:53   & 1.10     & 5120 & $\phi$=0.48    \\
V616 Mon  &  13/11/1997 & 14:56   & 1.23     & 5120 & $\phi$=0.74    \\
HD42606   &  13/11/1997 & 10:50   & 1.29     & 160  & K3V\\
\noalign{\smallskip}
BS5706    &  29/6/1995  & 06:23   & 1.12     & 96  & K0V \\
61 Cyg A  &  30/6/1997  & 15:49   & 1.16     & 96  & K5V \\
61 Cyg B  &  30/6/1997  & 15:46   & 1.16     & 96  & K7V \\
\noalign{\smallskip}
\hline
\end{tabular}
}
\end{table*}

\section{Observations and data reduction}

We obtained $K$-band (2.0--2.5$\mu$) spectra of V616 Mon using the
Cooled Grating Spectrometer (CGS4) on the United Kingdom 3.8-m Infrared
Telescope on Mauna Kea, during the night of 1997 November 13. The 40 l/mm
grating was used with the 150 mm camera and the 256$\times$256 pixel InSb
array. The bright star BS2233 was also observed through-out the night in
order to remove telluric atmospheric features. A journal of observations
is presented in Table 1. 

In order to minimise the effects of bad pixels, the standard procedure of
oversampling was used. The spectra were sampled over two pixels by
mechanically shifting the array in 0.5 pixel steps in the dispersion
direction, giving a full width half maximum resolution of 47 \AA\
($\sim$610 \kms at 2.31$\mu$). 
We employed the non-destructive readout mode of the
detector in order to reduce the readout noise. The slit width was 1.2
arcseconds which corresponds to 2 pixels on the detector. In order to
compensate for the fluctuating atmospheric emission lines we took
relatively short exposures and nodded the telescope primary so that the
object spectrum switched between two different spatial positions on the
detector. Throughout the observing run the slit orientation was north to
south in the spatial direction.

The CGS4 data reduction system performs the initial reduction of the 2D 
images.  These steps include the application of the bad pixel mask, bias 
and dark subtraction, flat field division, interlacing integrations taken 
at different detector positions, and co-adding and subtracting the nodded 
images (see Daly \& Beard 1994).  Extraction of the 1D spectra, 
wavelength calibration, and removal of the telluric atmospheric features 
was then performed using IRAF.  A more detailed description of the data 
reduction procedure is provided in Shahbaz et al. (1996).

First the individual spectra of V616 Mon were averaged using the 
variance in the spectrum as the weight. This resulted 
in a final summed spectrum which had a signal-to-noise ratio of $\sim$ 30.
We then cross correlated all the summed V616 Mon spectrum with the
template star spectrum in order to determine the velocity shift. Given
the poor resolution of the data this was not done on the individual
spectra of V616 Mon before they were averaged.
We applied the appropriate velocity shift and then
binned all the spectra onto a uniform velocity scale. Figure 1 shows the
summed spectrum of V616 Mon and the K3V template star spectrum,
normalised and shifted for clarity. The V616 Mon spectrum shows the CaI 
triplet and $^{12}$CO bands in absorption and doubled-peaked
Br$\gamma$ in emission. 

The peak separation of the Br$\gamma$ emission line arising from
the accretion disc is 1204$\pm$156km~s$^{-1}$. Note that this is
comparable with the peak-to-peak separation in the 
H$\alpha$ and H$\beta$ emission lines (MRW).
The equivalent width (EW) of Br$\gamma$ in the K3V star
spectrum is -1.6$\pm$0.3\AA\, whereas in the residual spectrum [i.e. the
spectrum of the accretion disc; see Figure 1 and section 3]
it is 14.6$\pm$1.3\AA. The $^{12}$CO bands
are the strongest features and will be used in the next
section to determine the fraction of light arising from the secondary
star.

\begin{figure*}
  \rotate[l]{\epsfxsize=450pt \epsfbox[0 0 700 800]{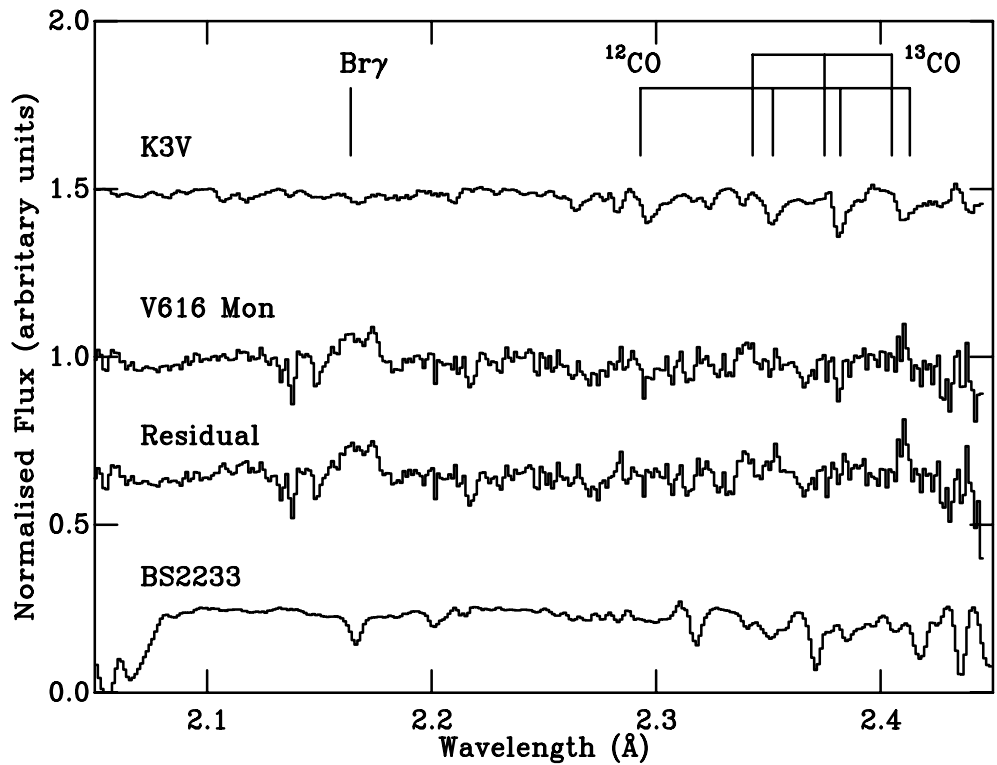}}
%  \picplace{8cm}
  \caption{ This figure shows the spectrum of V616 Mon and that of the
  template K3V star. The $^{12}$CO bandheads can clearly be seen.
  Also shown is the residual spectrum after optimally subtracting the
  template star from the V616 Mon spectrum. The lower-most spectrum is of
  an F6V star (BS2233) which indicates the location of telluric
  absorption features. All spectra have been normalised by dividing by a
  spline fit to the continuum. The stars indicate bad pixels. }
\end{figure*}

\section{The accretion disc contamination}

We first binned the V616 Mon and template star spectra 
onto a logarithmic wavelength scale using a $\sin\,x/x$ interpolation
scheme to minimize data smoothing (Stover et al. 1980). 
We then optimally subtracted the template star from the V616 Mon spectra
using the standard procedure (see MRW) as follows.
Taking the accretion disc contribution as a flat continuum, we fit the
V616 Mon spectra with this plus a variable fraction of the template star.
The fraction of light from the template star, ($f$), is adjusted to
minimise the residual scatter between the spectra. The scatter is
measured by carrying out the subtraction and then computing the \chis
between the residual spectrum and a smoothed version of the residual
spectrum. This is done to remove any large-scale structure. The \chis for
different fractions of the template star is computed and fitted with a
parabola, and the fraction at the minimum value of \chis is taken to be
the best fit. The above analysis was performed only on the 2.292$\mu$
$^{12}$CO(2,0) bandhead in the summed spectrum of V616 Mon. Other
bandheads were not used because of poor atmospheric subtraction and bad
pixels. Figure 1 shows the summed spectra of V616 Mon along with the
template K3V star. The residual light spectrum (i.e. the accretion
disc light) is also shown. 

In addition to using HD42606 (K3V) we
also used template stars of other spectral types
[HD42606 (K0V), HD42606 (K5V) and HD42606 (K7V)]
in order to determine the sensitivity of $f$ to the assumed spectral type.
We find that for the summed spectra of V616 Mon, the minimum value for
\rchis (1.01, a good fit) is obtained using the K3V secondary
star, which contributes 75$\pm$17 percent of the flux in the IR (the disc
fraction is 25 per cent). We also fitted the region used in the V616 Mon
spectrum for the optimal subtraction with a constant and obtained a much
worse fit with a \rchis of 2.7 thereby showing that the absorption feature
is real as the 99.9 per cent level.

The large uncertainties we obtain for the accretion disc light suggest
that our result is consistent with the secondary star contributing all
the light to the observed flux. In order to determine a lower limit to
the fraction of light from the secondary, we simulated a spectrum in
which {\it all} the light arises from the secondary star. This was done
using the K3V template star and adding noise to the data such that
the signal-to-noise was the same as the V616 Mon spectrum. We then
performed the optimal subtraction and determined the 2-$\sigma$ per cent lower
limit to the fraction of light arising from the secondary to be 73 per
cent. This value is comparable with the fraction of
light arising from the secondary star deduced from our spectrum of
V616 Mon, suggesting V616 Mon most probably
contains a companion star which contributes all the light in the IR.

\section{Discussion}

\subsection{Variability of the optical light curves}

The quiescent optical light curves of V616 Mon show a double-humped
ellipsoidal modulation due to the changing projected area of the Roche-lobe 
filling mass donor star. Maxima occur at quadrature, corresponding
to maximum projected area, and are hence expected to be symmetric.
Minima occur at orbital conjunctions; in general they have differing
depths due to limb- and gravity-darkening. The light from the
secondary star is diluted by the flux from the accretion disc. Also the
ellipsoidal light curve may be distorted by other variable contributions
to the total light; the bright spot associated with the impact of the
mass transfer stream on the edge of the disc (McClintock \& Remillard
1990); star spots on the secondary; and variable disc emission, including 
the superhump phenomenon (Warner 1995).

Recently Leibowitz, Herman \& Orio (1998) have collated the optical light
curves of V616 Mon obtained over the last 7 years. They find that the
depth of the maxima and minima of the light curves vary with time. They
observe a long term photometric behaviour of a few hundred days with a
peak to peak amplitude of 0.3 mag. They suggest that the minimum in the
light curve corresponding to when the red dwarf lies between the observer
and the compact object (orbital phase 0.0) changes depth with time. Since
in the standard precessing disc model one does not expect this minimum to
change depth significantly (as at this phase the secondary star is least
affected by X-ray illumination) they conclude that they cannot
explain the variations in terms of a simple geometrical precessing
accretion disc model (Heemskerk \& van Paradijs 1989).

However, it should be noted that the main thrust of their conclusion lies
in the interpretation of the varying component in the light curves.
In the precessing disc model the depth
of either minimum can vary with time, depending on the tilt of the
accretion disc (Heemskerk and van Paradijs 1989). Also they have assumed
that the maximum in the light curve (orbital phase 0.75) with respect to
which the minima is measured remains constant. This is probably not the
case as the observed optical light near this orbital phase is
contaminated by the bright spot. Optical spectroscopy of V616 Mon shows
the presence of a bright spot (MRW). Bright spots have been
seen in the optical light curves of cataclysmic variables with 
similar mass ratios such as Z Cha and OY Car (Wood et al. 1986, 1989) where it is
observed between orbital phase 0.6 and 1.1. Variations in the optical
flux emitted by the bright spot as the result of clumpy mass transfer
from the secondary star could easily give rise to variability in the
optical flux observed at this orbital phase.

In the IR the effects of the variability discussed above are much less
(SNC). The IR light curve of V616 Mon shows
equal maxima; this is what is expected if the IR variations are due solely to
the ellipsoidal modulation of the secondary star.

\begin{table}
\centering
\small{
\caption{Optimal Subtraction of the Companion Star}
\begin{tabular}{lccc}\hline
\hline\noalign{\smallskip}
Star     & Sp. Type     & $\chi^{2}$ (DOF=13)  &   $f$ \\ \hline
BS5706   & K0V 	& 23.7                 &  35 $\pm$ 12 \\
HD42606  & K3V 	& 13.2                 &  75 $\pm$ 17 \\ 
61 Cyg A & K5V 	& 31.7                 &  13 $\pm$ 13 \\ 
61 Cug B & K7V 	& 32.6                 &   5 $\pm$ 12 \\ \hline
\noalign{\smallskip}
\hline
\end{tabular}
}
\end{table}

\subsection{The Brackett-$\gamma$ emission line}

Various authors have pointed out that the double-peaked emission line
profiles can be interpreted as arising from an accretion disk viewed at
high inclination. However, it should be noted that a double-peaked
emission line profile can also arise from a system with an inclination as
low as 15$^\circ$ (see the models of Horne \& Marsh 1986). Assuming that
the double-peaked lines arise entirely from the disk, we can estimate the
binary inclination of V616 Mon by measuring the separation of the
Br$\gamma$ emission-line peaks.

The Keplerian velocity of the outer edge of the disk ($v_{d}$) is given
by $v_{d}=(G M_{1}/R_{d})^{0.5}$. Combining this with Kepler's third law,
Paczynski's (1971) formula for the Roche lobe radius, and using the fact
that the accretion disk fills 50 per cent of the compact object's Roche-lobe
(MRW), gives $v_{d}=998 (M_{1}/P_{hr})^{1/3}$ km s$^{-1}$,
where $M_{1}$ is the mass of the black hole (in solar masses), and
$P_{hr}$ is the orbital period (in hours). The separation of the
Br$\gamma$ emission-line peaks measured from the summed spectrum of V616
Mon is 1204 km~s$^{-1}$, implying a projected velocity of the outer edge
of the accretion disk of $v_{d} \sin i \sim$ 602 km~s$^{-1}$. Using the
above formula with $P_{hr}$=7.75 hrs (MRW) and $M_{1} \sim
10 M_{\odot}$ (SNC) gives a $i\sim 34^{\circ}$, which 
agrees well with that obtained by SNC.

In cataclysmic variables there is observational evidence that the
accretion disc contamination in the $K$-band is significant. The eclipse
light curves of the dwarf nova OY Car, for example show that during
quiescence the accretion disc can contribute about 30 percent of the flux
at 2.2$\mu$ (Sherrington et al. 1982). IR spectra of cataclysmic
variables also show emission lines arising from the optically thin gas in
the accretion disc (Ramseyer et al. 1993; Dhillon \& Marsh 1995), such as
HeI (2.0587$\mu$) and Br$\gamma$ (2.1655$\mu$). In contrast, the
X-ray transients V404 Cyg (Shahbaz et al. 1996) and V616 Mon show only
Br$\gamma$ in emission. It should also be noted that the mass accretion
rate during quiescence in the X-ray transients is a factor of 10 lower
than that in dwarf novae.

One expects the EW of the emission lines arising from the accretion disc
to decrease as the orbital period of the binary increases. This is simply
because larger systems will have larger, cooler accretion discs. If one
looks at the H$\alpha$ EW in the SXTs, then one can find some evidence
for this type of correlation; for Nova Per 1992 the H$\alpha$ EW is
205\AA, whereas for the larger systems such as Nova Mus 1991 and V404 Cyg
it is 50\AA\ and 40 \AA\ rrespectively. Also note that in all the SXTs the
disc contamination near H$\alpha$ is in the range 6-16 percent, i.e. it
is small despite the large H$\alpha$ EW. In V404 Cyg the Br-$\gamma$ EW
of the accretion disc is 2.7\AA. One expects the Br-$\gamma$ EW of the
accretion disc in the much shorter system V616 Mon to be higher; this is
what is observed (see section 2).

\subsection{The effect on the mass of the black hole}

SNC obtained an IR light curve of V616 Mon which showed a double humped
feature characteristic of the ellipsoidal variations of the secondary
star. They modelled these variations assuming all the IR flux was arising
from the secondary star, and determined the most probable mass of the
compact object to be 10 \Msun. Justification for this assumption comes
from the analysis of the IR light curve of the transient Cen X--4
(Shahbaz et al. 1993). The mass of the compact object in Cen X--4 is
consistent with that of a canonical neutron star; which the compact
object must be because of the type I X-ray bursts observed during
outburst (Matsuoka et al. 1980). This provides indirect evidence that the
contribution of the accretion disc to the observed IR flux is very small,
at least from Cen X--4. This may also be the case for V616 Mon; an upper
limit to the accretion disc contribution to the IR flux being 27 percent
(section 3).

The effects of any accretion disc contamination to the
observed IR flux will be to dilute the actual ellipsoidal modulation, 
making the observed modulation smaller than the {\it true}
value. Since the amplitude of the ellipsoidal modulation is
correlated with the binary inclination (large amplitudes imply a high
binary inclination), this means that modelling a contaminated light curve
will underestimate $i$.

We have modelled the amplitude of the ellipsoidal variations as a function of
$i$. We used the same parameters as SNC:
T$_{eff}$=4000 K, $q$=14.9 (MRW), a gravity darkening exponent of 0.08
(Lucy 1967), and the limb darkening coefficient from Al-Naimiy (1978).
Figure 2 shows the effect of differing amounts of contamination in the IR
light curves on the binary inclination. If we take the 2-$\sigma$ 
limit to the disc contamination of 27 percent, we find that $i$
increases by 7 degrees and the mass
of the black hole decreases by 3.6 \Msun (2-$\sigma$ limit).
Note that this extreme value for $i$ is still lower
than that obtained by Haswell et al. (1993), and the implied mass of the
compact object ($\sim$ 6.4 \Msun) still substantially exceeds the
canonical maximum mass of a neutron star (3.2 \Msun; 
Rhoades \& Ruffini 1974).

\begin{figure*}
  \rotate[l]{\epsfxsize=450pt \epsfbox[0 0 700 800]{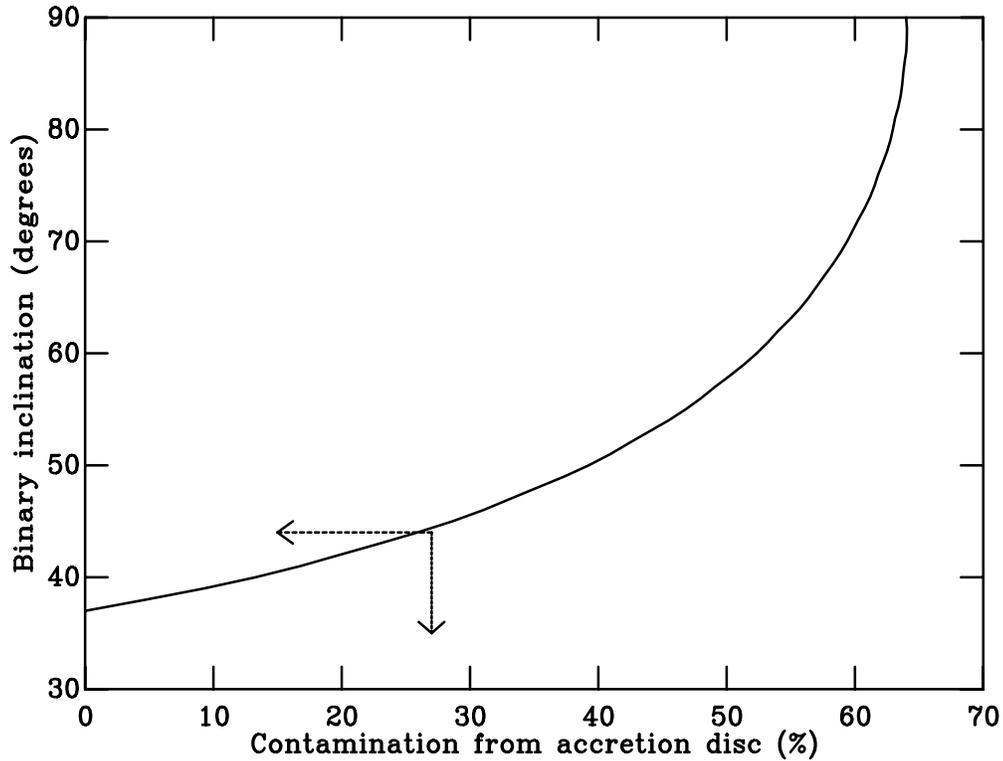}}
%  \picplace{8cm}
  \caption{ The effect of the accretion disc contamination on the value
  for the binary inclination, derived by modelling the $K$-band
  ellipsoidal variations of the secondary star (SNC). The 2-$\sigma$ 
  upper limits to the accretion disc contamination (0--27 percent) and
  binary inclination (37$^{\circ}$--43$^{\circ}$) are shown. }
\end{figure*}

\section{Conclusion}

We have determined the contamination of the observed IR flux by the
accretion disk in V616 Mon. We obtained IR spectra of V616 Mon which when
optimally subtracted from the secondary star spectrum (K3V star)
allows us to determine the 2-$\sigma$ upper limit to the accretion disc
contribution to be less than 27 percent. We find that this only increases
the determination of the binary inclination by less than 7 degrees; the
mass of the black hole is decreased by less than 3.6 \Msun.

\section*{Acknowledgements}

We would like to thank T. Geballe for useful discussions.
The data reduction was carried out using the IRAF and ARK 
software packages. The United Kingdom Infrared Telescope is operated by the
Royal Observatories on behalf of the UK Particle Physics and Astronomy
Research Council.

\end{document}